\documentclass[%
 reprint,
superscriptaddress,
nofootinbib,
 amsmath,amssymb,
 aps,
]{revtex4-1}
\usepackage[normalem]{ulem}
\usepackage{color}
\usepackage[dvipsnames]{xcolor}
\usepackage{slashed}
\usepackage{graphicx}
\usepackage{dcolumn}
\usepackage{bm}

\graphicspath{ {images/} }

\begin{document}

\preprint{APS/123-QED}

\title{Constraining free parameters of a color superconducting non-local Nambu-Jona-Lasinio model using Bayesian analysis of neutron stars mass and radius measurements}

\author{M. Shahrbaf}
\affiliation{%
Institute for Theoretical Physics, University of Wroclaw, Max Born Pl. 9, 50-204, Wroclaw, Poland
}%
\author{S. Anti\'{c}}
\thanks{Guest scientist}

\affiliation{%
Institute for Theoretical Physics, University of Wroclaw, Max Born Pl. 9, 50-204, Wroclaw, Poland
}%
\affiliation{%
$^\ast$ Physics Department, Faculty of Science, University of Zagreb, Bijenička cesta 32, 10000 Zagreb, Croatia 
}%

\author{A. Ayriyan}
\affiliation{%
Meshcheryakov Laboratory of Information Technologies, Joint Institute for Nuclear Research, Joliot-Curie str. 6, 141980 Dubna, Russia
}%
\affiliation{%
Dubna State University, Universitetskaya Str. 19, Dubna, 141980, Russia
}%
\affiliation{%
IT and Computing Division, A. Alikhanyan National Laboratory, Alikhanian Brothers Str. 2, Yerevan, 0036, Armenia
}%

\author{D. Blaschke}
\affiliation{%
Institute for Theoretical Physics, University of Wroclaw, Max Born Pl. 9, 50-204, Wroclaw, Poland
}%


\author{A. G. Grunfeld}
\affiliation{%
CONICET, Godoy Cruz 2290, Buenos Aires, Argentina
}%
\affiliation{%
Departamento de F\'\i sica, Comisi\'on Nacional de Energ\'{\i}a At\'omica, Av. Libertador 8250, (1429) Buenos Aires, Argentina
}%

\date{\today}

\begin{abstract}
We provide a systematic study of hybrid neutron star equations of state (EoS) consisting of a relativistic density functional for the hadronic phase and a covariant nonlocal Nambu--Jona-Lasinio (nlNJL) model to describe the color superconducting quark matter phase. 
Changing the values of the two free parameters, the dimensionless vector and diquark coupling strengths $\eta_V$ and $\eta_D$ results in a set of EoS with varying stiffness and deconfinement onset. 
The favorable parameters are obtained from a systematic Bayesian analysis for which 
the multi-messenger constraint on the neutron star radius at $1.4~$M$_\odot$ and the combined mass-radius
constraint for PSR J0740+6620 from NICER experiment are used as the constraints. 
Additionally, the transition from hadronic matter to deconfined quark matter is constrained to occur above nuclear saturation density. 
Hybrid stars modeled with these favorable parameters are compatible with the NICER results for the radius of the highest known mass neutron star, PSR J0740+6620.
Three new observations interesting for neutron star phenomenology are reported:
1) We show that the constant sound speed (CSS) EoS provides an excellent fit to that of the nlNJL model which implies the squared speed of sound at high densities to be about $0.5$ for the optimized parameters;
2) we give a simple functional form for the mapping between the parameter spaces of these two models valid for the whole range of relevant chemical potentials and
3) we observe that the special point property of hybrid EoS based on CSS quark matter generalizes to a set of lines consisting of special points when two EoS parameters are varied instead of one. A lower limit for the maximum mass of hybrid stars as a function of the vector coupling strength is obtained.

\end{abstract}

\maketitle


\section{Introduction}
{The last few years have been a brilliant time for astrophysics with various observations made on compact stars in isolation and in binaries yielding their mass, radius and tidal deformability \cite{riley2021nicer, miller2021radius, dietrich2020multimessenger, abbott2018gw170817, fonseca2021refined, miller2019psr}. These observations provide important data constraining the equation of state (EoS) of dense matter, the joining element between astrophysics, nuclear and particle physics. In particular, the most recent results of NICER for the radius of the most massive observed pulsar, PSR J0740+6620 \cite{riley2021nicer, miller2021radius}, introduce a challenge for hadronic matter EoSs which have to result in a maximum neutron star (NS) mass above $2\textmd{M}_{\odot}$ and a relatively large radius above $12$ km at this high mass. 
At the same time, the EoS should be soft enough around $1.4\textmd{M}_{\odot}$ to fulfill the tidal deformability constraint from the binary neutron star merger GW170817. 
This challenge can be faced by constructing a phase transition from hadronic matter to deconfined quark matter that occurs below $1.4\textmd{M}_{\odot}$. 
Although a transition to quark matter softens the EoS, the quark EoS should be rather stiff to fulfill the observational constraints.}
Recent investigations of neutron star (NS) properties support the idea that deconfined quark matter builds the cores not only of the heaviest known NS \cite{Annala:2019puf}, but even of all NS in the presently observed mass range \cite{Blaschke:2020qqj}. In order to draw such conclusions, the equation of state for the quark matter phase in NS is needed. 

The best candidates to provide a reliable quark matter EoS under NS conditions are effective models that share key features with low-energy QCD, such as dynamical chiral symmetry breaking and the resulting low-energy theorems as the Gell-Mann--Oakes--Renner and Goldberger--Treiman relations.  
The most popular chiral effective model that reasonably fulfills these requirements is the Nambu-Jona-Lasinio (NJL) model~\cite{Nambu:1961tp,Nambu:1961fr}, that was developed in order to understand the generation of a mass gap for fermions (nucleons) in analogy to the energy gap in the BCS model of electronic superconductivity, based  on local four-fermion interactions. 
With the introduction of quarks as the fundamental fermionic degrees of freedom of strongly interacting matter, the model has been reformulated on that level of description, see \cite{Eguchi:1976iz,Volkov:1984kq,Vogl:1989ea,Klimt:1989pm,Klevansky:1992qe,Ebert:1994mf,Hatsuda:1994pi} for early reviews and \cite{Buballa:2003qv} for the NJL model analysis of dense quark matter.   

The local NJL-type models have a major caveat: the absence of confinement. A local four-fermion coupling is incompatible with the strong coupling at large-distances 
that is phenomenologically established by hadron spectroscopy and lattice QCD simulations of the free (potential) energy between static (heavy) color charges
that exhibits a funnel-shaped interquark potential.
One way to bring effective low-energy QCD models closer to capturing also the confining nature of the strong interaction at large distances is the generalization of the local four-fermion coupling to a nonlocal four-point function.
A first step in this direction has been done with so-called "bilocal QCD" models \cite{Kleinert:1976ds,Ebert:1976rh}
which could successfully address low-energy QCD and hadron phenomenology on the basis of a nonperturbative model gluon propagator \cite{Cahill:1985mh} or a relativistic generalization of confining potential models \cite{Pervushin:1989ee}.

The next step towards the modern formulation of the nonlocal chiral quark model (in the following denoted as "nlNJL") was to generalize the interaction model from a two-point function to a four-point function, following two schemes. 
The first one defines the nonlocal interaction similar to the relativistic S-matrix with its dependence on the kinematic Mandelstam variables and assumes separability (a product ansatz) of pairwise relative momenta \cite{Schmidt:1994di}.  
This scheme has later been denoted as one-gluon exchange (OGE) model.
For the model formfactors the ansatz has been made of instantaneous (energy-independent) functions that depend on the relative three-momenta only. In this way, the evaluation of thermodynamic properties of the model within the imaginary time formalism became straightforward because the Matsubara summations could still be performed analytically as in the original NJL model
which was recovered for a step function as the formfactor. 
Using smooth functions as formfactors of the nonlocality resulted in a lowering of the psudocritical temperature of the chiral restoration, closer to the results from lattice QCD \cite{Schmidt:1994di}. 
The second scheme, denoted as instanton-liquid model (ILM), was built using a simple product ansatz for the formfactors that were attached to each of the four fermion "legs"
\cite{Bowler:1994ir}.
A comprehensive overview over the nonlocal chiral quark models in this early era has been given by Georges Ripka in his book 
\cite{Ripka:1997zb}, where formal details and more references can be found.

With the extension to rank-2 separable gluon propagator models, 
and using formfactors depending on the 4-momentum, a close correspondence to the Dyson-Schwinger equation approach could be established \cite{Burden:1996xs}. The finite-temperature extension led to a dramatic reduction of the chiral restoration temperature 
\cite{Blaschke:2000gd} which recently has been found in lattice QCD simulations as $T_c^0=132^{+3}_{-6}$ MeV in the chiral limit
\cite{Ding:2019prx}.
When coupling the nonlocal chiral quark dynamics to a gluon background field within the Polyakov-loop extension of the model
(see \cite{Ratti:2005jh} for the details in case of the local NJL model, where besides adding gluonic degrees of freedom at finite temperatures confinement is mimicked), 
acceptable values and systematics for the pseudocritical temperature of QCD could be obtained \cite{Horvatic:2010md,Radzhabov:2010dd} and thermodynamic instabilities associated with complex-conjugated mass poles could be largely cured \cite{Benic:2012ec}. 

Another advantage of using covariant formfactors in the nlNJL model is that the momentum dependence of the quark mass as well as the wave function renormalization factor of the quark propagator can be described in accordance with lattice QCD simulations \cite{Noguera:2008cm}.
Since the nonperturbative low-energy interaction of the nlNJL model can be calibrated using lattice QCD simulations, one has a serious basis to extend studies of thermodynamic properties from the temperature axis into the whole phase diagram including high baryochemical potentials where no lattice QCD simulations are available because of the severe sign problem. 
Using the nlNJL model with realistic formfactors, the position of the critical endpoint in the phase diagram has been obtained for temperatures below 130 MeV \cite{Contrera:2012wj,Contrera:2016rqj}, a prediction that can be investigated in the upcoming heavy-ion collision experiments with collision energies $\sqrt{s_{NN}}\sim 3 - 6$ GeV, e.g., at the NICA facility of JINR Dubna, or in the fixed-target experiment at RHIC, see Fig. 14 of Ref.~\cite{Senger:2021dot}.  
A recent detailed review on nlNJL models and their applications to studies of matter under extreme conditions is given in \cite{Dumm:2021vop}. 

In the present work we want to study the EoS of quark matter in NSs
at vanishing temperature. 
A first application of the covariant nlNJL model has been given in 
\cite{Gocke:2001ri}, where a simple Gaussian formfactor was used and only the scalar-pseudoscalar meson interaction channel has been included. It could be shown that such a model with two quark flavors is equivalent to a thermodynamic bag model with a  bag pressure $B=81.3$ MeV/fm$^3$. The strange quark flavor appeared sequentially at a higher density and rendered the corresponding neutron star unstable against gravitational collapse for masses 
above $1.62~\textmd{M}_\odot$.  
In order to describe stars with masses above $2~\textmd{M}_\odot$ as required by recent neutron star observations \cite{fonseca2021refined}, one has to include a repulsive vector meson interaction channel\footnote{We note that in the local limit such a model has been introduced as "vBag model" \cite{Klahn:2015mfa}.}.
In hybrid star models the transition to such a quark matter phase
occurs only at rather high densities, close to the maximum mass of the sequence. 
As it is known from studies using the local NJL model, the simultaneous addition of a scalar diquark interaction channel leads to the effect of diquark condensation (color superconductivity) which pushes the onset of quark deconfinement to lower densities and results in quark matter cores for neutron stars with typical masses of $\sim 1.4 \textmd{M}_\odot$ \cite{Klahn:2006iw,Klahn:2013kga}.
The situation is similar for the covariant nlNJL model where increasing the diquark coupling results in a lowering of the onset density for the chiral restoration transition \cite{GomezDumm:2005hy}. 
Correspondingly, a phenomenologically satisfactory situation occurs for the hybrid neutron star EoS when a repulsive vector and diquark interaction channel are included to the covariant nlNJL model \cite{Blaschke:2007ri} since it develops an early deconfinement transition to a stiff color superconducting quark matter phase.

Consequently, in this study we will use the covariant nlNJL model in the OGE scheme that is parameterized with the dimensionless vector and diquark coupling strengths ($\eta_V$ and $\eta_D$), given as ratios of the vector and diquark couplings, $G_V$ and $G_D$, to the scalar coupling constant $G_S$, respectively. The values of these input parameters, however, are not known from first principles. 
{In Ref.~ \cite{Contrera:2022tqh} a similar analysis is performed in the frame of the instantaneous nlNJL interaction model with three-dimensional (3D) form factor in the momentum space, to study a family of hybrid EOS for compact stars.}

We will perform the Bayesian analyses based on the observational constraints of NS to investigate the most likely values for these two parameters which results in the most likely EoS. 
From the Bayesian analysis in \cite{Blaschke:2020qqj}, it is concluded the phase transition onset most likely occurs in the center of neutron stars with masses around
$1 \textmd{M}_{\odot}$ which is in agreement with the observed compactness \cite{capano2020stringent}. Therefore, we use a special parameterization of generalized relativistic density
functional (GRDF) model that is called DD2p00 ( \cite{Typel:2009sy}, without excluded volume) and DD2p40 (\cite{Typel:2009sy, Typel:2016srf}, with excluded volume) with baryon-meson couplings that
depend on the total baryon density of the system. It is a stiff hadronic matter EoS which supports having a phase transition onset at low densities. It is worth mentioning that the Bayesian analysis of multimessenger M-R data with interpolated hybrid EoS has been also investigated recently in \cite{Ayriyan:2021prr} when the hadronic EoS has been fixed to be APR \cite{Akmal:1998cf}. In the present work, we investigate the opted parameters of nlNJL model via a Bayesian analysis based on a Maxwell construction for the phase transition from DD2 model to nlNJL one. It is worth mentioning that compared to APR EoS, DD2 has been extended to include hyperons and fulfills the maximum mass constraint of NS \cite{Shahrbaf:2022upc}.

While the original version of the nlNJL model is formulated with constant, density-independent coefficients, 
in Refs.~\cite{Alvarez-Castillo:2018pve,Ayriyan:2018blj} 
density-dependent coefficients have been introduced in such a way that the results of a recent relativistic density-functional approach to quark matter \cite{Kaltenborn:2017hus} could be reproduced. The applicability of both versions of nlNJL model in constructing the phase transition from hypernuclear matter to deconfined quark matter has also been investigated \cite{Shahrbaf:2019vtf, Shahrbaf:2020uau}.

The microscopic approach on the basis of the nlNJL model has the advantage that it allows to determine the ranges of the parameters ($\eta_V$ and $\eta_D$) in the Lagrangian of an effective low-energy QCD model 
using the spectrum of observable NS properties. For details, see Ref.~\cite{ayriyan2021bayesian}. 
This strategy, however, involves time-consuming numerical routines for solving the self-consistent nonlocal mean field equations together with an extrapolation procedure at high densities. 
For this reason, we explore the possibility to mimic the nlNJL results with a simpler approach, the constant-sound-speed (CSS) EoS model \cite{Alford:2013aca,Zdunik:2012dj}. 

The CSS approach is widely used in the literature, in particular, for the classification \cite{Alford:2013aca} and systematics \cite{Blaschke:2020vuy, Miao:2020yjk} of hybrid neutron stars. Among the applications of the CSS model is also the investigation of the third and fourth families of compact stars for which stable branches have been verified as well  \cite{Paschalidis:2017qmb, Alford:2017qgh,Li:2019fqe}. The work of Ref.~\cite{Zdunik:2012dj} demonstrates the possibility for NJL model-based approaches to color-superconducting cold quark matter to be well approximated by CSS parameterization. It was shown that the EoS for quark matter developed for the nonlocal separable NJL model with formfactors depending on the three-momentum in \cite{Blaschke:2005uj,Blaschke:2010vd} can be well fitted with the CSS model, see also \cite{Contrera:2022tqh}.

It is worth mentioning that the CSS extrapolation becomes necessary for nlNJL models of certain values of $\eta_V$ and $\eta_D$ parameters due to the limitation of its covariant formfactor realization \cite{GomezDumm:2005hy} to chemical potentials up to $\sim 1600$ MeV. 
However, we would like to map the nlNJL EoS to the CSS EoS for the whole range of chemical potentials in the quark matter phase, and not only for $\mu_B > 1600$ MeV, as it was previously done \cite{ayriyan2021bayesian}. 
This mapping would enable a replacement of the complicated quark matter EoS by a simple model which gives the EoS of quark matter not only in the two-flavor color-superconducting (2SC) phase but also at higher densities for the color-flavor-locked (CFL) phase. 

{Moreover, another significant aspects of employing the CSS EoS at high densities is appearance of the special point on the Mass-Radius (M-R) diagram of the hybrid stars. An analytical study has been performed employing a CSS EoS that explains the existence of a very small region on the 
M-R plot of hybrid stars where all of the lines representing the sequences of models with different constant values of the bag pressure intersect \cite{Yudin:2014mla}.}

The main idea of this work is to provide a 
systematic study on the parameters of nlNJL model that vary in the range $0.10 < \eta_V < 0.20$ and $0.70 < \eta_D < 0.80$ based on Bayesian analysis which are performed with respect to the modern mass and radius constraints of NS. We find the most probable parameters of quark matter model in hybrid EoS using a Maxwell construction which is very well compatible with the observed constraints and in particular, with the most recent results of NICER for the radius of NS. 

Moreover, a simple functional form is found in this work that enables a mapping between the two parameter spaces, the nlNJL model parameters $\eta_V$ and $\eta_D$ and the parameters of the CSS model: the slope parameter $A$, the squared speed of sound $c_s^2$ and the bag pressure $B$. 
With the simplified description of the quark phase in hybrid NS, its EoS would become easier to handle and would at the same time have strong microphysical justification. 
{Following the fact that the nlNJL 
EoS appears to be isomorph to a CSS parameterization for the high-density phase, we show that the special point properties discussed in \cite{Yudin:2014mla} generalizes to a set of lines consisting of special points, when two parameters ($\eta_V$ and $\eta_D$) are changed instead of one parameter (bag pressure).}

The structure of the present paper is as follows: in Sec. \ref{sec:EOSmodels} we introduce the formulations of both nlNJL and CSS models and their parameters. 
The results for mass, radius and tidal deformability of obtained hybrid stars are shown in Sec. \ref{sec:Results}. The results of the Bayesian analysis with the astrophysical inputs for these analysis are presented in Sec. \ref{sec:Bayasian}.  
Finally, we provide the conclusion of this study in Sec. \ref{sec:conslusion}. The functional dependence between the two parameter sets of the CSS and nlNJL models is found and the parameter mapping between the two models is discussed in Appendix A. 
We present our analysis on special points in the mass-radius diagram in Appendix B. Moreover, in Appendix C, a phenomenological EoS is introduced to be investigated how well nlNJL EoS is fit to it compared to CSS parameterization.

\section{Equation of state models for the quark matter phase of a neutron star} \label{sec:EOSmodels}
\subsection{Generalized nlNJL model with $\mathbf{\eta_V}$ and $\mathbf{\eta_D}$ parameters }

For the microphysical description of the quark matter phase we consider a chiral quark model that includes
nonlocal separable interactions and can be considered as a nonlocal extension of the NJL model (nlNJL). 
We employ the two-flavor $SU(2)_f$ model, developed in Refs.~\cite{Schmidt:1994di,GomezDumm:2005hy,Blaschke:2007ri}, that is described by the Lagrangian
\begin{eqnarray}
\mathcal{L} &=& \bar\psi\left(-i\slashed\partial+m_c\right)\psi - \frac{G_S}{2} j_S^f j_S^f 
- \frac{G_D}{2} [j_D^a]^\dagger j_D^a
+ \frac{G_V}{2} j_V^\mu j_V^\mu~,\nonumber\\
\end{eqnarray}
with the nonlocal generalizations of the quark currents
\begin{eqnarray}
j_S^f(x)&=& \int d^4z g(z) \bar\psi(x+\frac{z}{2})\Gamma^f \psi(x-\frac{z}{2})~,
\\
j_D^a(x)&=& \int d^4z g(z) \bar\psi_C(x+\frac{z}{2})i\gamma_5 \tau_2 \lambda^a \psi(x-\frac{z}{2})~,
\\
j_V^\mu(x)&=& \int d^4z g(z) \bar\psi(x+\frac{z}{2})i\gamma_\mu
\psi(x-\frac{z}{2})~,
\end{eqnarray}
in the scalar meson, scalar diquark and vector meson channels, respectively. 
The grand canonical partition function of the quark matter system, 
\begin{equation}
    \mathcal{Z} = \int \mathcal{D}\bar{\psi} \mathcal{D}\psi
    \exp\left\{-\int_0^\beta d\tau \int d^3x \left[ \mathcal{L} - \mu \bar{\psi} \gamma_0 \psi\right] \right\}
    ~,
\end{equation}
after bosonization by the Hubbard-Stratonovich transformation, can be evaluated in the mean field approximation (MFA) with the result for the thermodynamical potential
\begin{eqnarray}
\Omega^{\rm MFA}  &=&  - T \ln \mathcal{Z}^{\rm MFA}\\
&=&\frac{ \bar
\sigma^2 }{2 G} + \frac{ {\bar \Delta}^2}{2 H} 
- \frac{\bar \omega^2}{2 G_V} \nonumber\\
& &-\frac{1}{2} \int \frac{d^4 p}{(2\pi)^4} \ \ln
\mbox{det} \left[ \ S^{-1}(\bar \sigma ,\bar \Delta, \bar \omega,
\mu_{fc}) \right] ~,
\label{eq:mfaqmtp}
\end{eqnarray}
see Refs.~\cite{GomezDumm:2005hy,Blaschke:2007ri} for details.

The inclusion of the scalar diquark channel together with the repulsive vector interaction channel, plays an important role in the phenomenology of hybrid EoS of compact stars.

The diquark condensate gives rise to color superconductivity (2SC) and is responsible for lowering the onset of the phase transition from the phase with broken chiral symmetry to the 2SC phase. The vector interaction induces a stiffening behaviour in the EoS, that is essential to reach compact stars masses above $2 \textmd{M}_{\odot}$.
Systematic investigation of hybrid NS properties reveals
\cite{ayriyan2021bayesian,Klahn:2013kga} that phenomenological constraints from mass and radius measurements are optimally fulfilled when an increase in the diquark coupling is accompanied by a simultaneous increase in the vector coupling.  

The model includes three input parameters: $m_c$ (current quark mass), $p_0$ (effective momentum scale) and $G_S$ (coupling constant). They are determined as to reproduce the pion mass and decay constant as well as the chiral condensate in the vacuum, at vanishing temperature and densities. The two remaining coupling constants $G_S$ and $G_V$ are driving the terms that, after bosonization, give rise to the color superconducting gap and the vector meson mean field. 
The dimensionless ratios $\eta_D = G_D/G_S$ and $\eta_V = G_V/G_S$ are free parameters. 
From a Fierz rearrangement of the OGE interactions one obtains $\eta_D =3/4$ and $\eta_V = 1/2$ that could serve as an orientation for the values of these parameters in the vacuum. There is no precise derivation of effective couplings from 
QCD, as we consider here the strongly nonperturbative low-energy regime. Moreover, one has to expect that these couplings could be subject to a medium dependence.  
However, $\eta_D$ values larger than  
$\eta_D^{*} = (3/2)m/(m - m_c)$ 
may lead to color symmetry breaking in the vacuum \cite{Zablocki:2009ds} (where $m$ stands for the dressed mass and $m_c$ for the current quark mass). 

In the present work we consider a window of values for $\eta_D$ and $\eta_V$ that was also explored in previous works Ref.~\cite{Alvarez-Castillo:2018pve,Klahn:2013kga, Ayriyan:2021prr}.

The mean field values $\bar \sigma$, $\bar \Delta$ and $\bar \omega$ are obtained from the coupled equations
\begin{equation}
\frac{ \partial \Omega^{\rm MFA}}{\partial\bar \sigma} = 0, \quad 
\frac{ \partial \Omega^{\rm MFA}}{\partial\bar \Delta} = 0, \quad 
\frac{\partial \Omega^{\rm MFA}}{\partial\bar \omega} = 0.
\label{eq:gapeq}
\end{equation}
As we intend to describe the behaviour of quark matter in the cores of NSs, we have to take into account the presence of leptons (electrons and muons) which we include into the thermodynamic potential as free relativistic Fermi gases. 
In addition, we have to consider that the stellar matter satisfies the following conditions: equilibrium under weak interactions (chemical equilibrium) as well as color and electric charge neutrality. 
As a consequence, it can be seen that the six different chemical potentials $\mu_{fc}$  (depending on the two quark flavors $u$ and $d$ and quark colors $r,g$ and $b$) in 
Eq.~\eqref{eq:mfaqmtp} are not independent from each other and can be written in terms of three independent quantities: the baryonic chemical potential $\mu$, the electron chemical potential $\mu_e$ and a color chemical potential $\mu_8$. 
Basically, for each value of $\mu$ we solve self-consistently the gap equations (\ref{eq:gapeq}), complemented with  the  conditions for $\beta$-equilibrium and electric charge and color charge neutrality (details of the calculation can be found in the Appendix of Ref.~\cite{Blaschke:2007ri}).

In the present work, we consider a Gaussian form factor $g(p)=\exp (-p^2/p_0^2)$ in Euclidean 4-momentum space. The fixed input parameters of the quark model considered here are $m_c=5.4869$ MeV, $p_0=782.16$ MeV and $G_S p_0^2 = 19.804$.

\subsection{Constant speed of sound formulation}

The CSS EoS at zero temperature can be written in the form 
\cite{Alford:2013aca,Blaschke:2020vuy}
\begin{equation}
\label{eq:CSS_EoS}
    P(\mu) = A \left( \frac{\mu}{\mu_x}\right)^{1+\beta} - B ,
\end{equation}
where $\mu_x = 1$ GeV defines the scale for chemical potential, $A$ is a slope parameter in the units of the pressure, $B$ is the bag pressure and $\beta = 1/c_s^2$ is a parameter related to the squared speed of sound $c_s^2 = dP/d\varepsilon$.
The pressure as a function of the chemical potential is a thermodynamical potential from which other EoS can be obtained by derivation.
For instance, the baryon density reads
\begin{eqnarray}
n_B(\mu) = \frac{dP(\mu)}{d \mu} = A \frac{1+\beta}{\mu_x} \left(\frac{\mu}{\mu_x} \right)^{\beta}~,
\end{eqnarray}
and the energy density is given by
\begin{eqnarray}
\label{eq:epsilon}
\varepsilon(\mu) = \mu \frac{dP(\mu)}{d \mu} - P(\mu)
= A \beta \left(\frac{\mu}{\mu_x}\right)^{1+\beta} + B.
\end{eqnarray}
Using the definition of the pressure (\ref{eq:CSS_EoS}), the 
energy density (\ref{eq:epsilon}) can be rewritten as 
\begin{eqnarray}
\label{eq:epsilon-p}
\varepsilon(P) =  \beta P + (1+\beta) B~,
\end{eqnarray}
which directly reveals that the squared sound speed is
$c_s^2=1/\beta = {\rm const}$, since $\beta = {\rm const}$. 

The speed of sound determines the stiffness of the EoS, which has to be large enough to allow for the maximum neutron star mass to fulfil the observational lower bound of $2.01~\textmd{M}_\odot$ from the Shapiro-delay based mass measurement on PSR J0740+6620 \cite{fonseca2021refined} (a recent upgrade of the former mass measurement \cite{Cromartie:2019kug}) at the $1\sigma$ level. 
The prefactor $A$ changes the slope of the $P(\mu)$ curve and
has thus also an effect on the stiffness of the EoS: lowering the value of $A$ increases the stiffness.
The effective (negative) bag pressure $B$ realises quark confinement at low densities in quark matter EoS because it makes sure that any small but positive pressure of a hadronic phase would be preferable this region. 
The parameters $A$, $B$ and $c_s^2$ are free parameters which define the behaviour of the quark matter EoS.

It is worth mentioning that for a large class of quark matter models, including the standard NJL model \cite{Agrawal:2010er, Bonanno:2011ch} or its nonlocal generalization with instantaneous, three-momentum dependent formfactors 
\cite{Lastowiecki:2011hh}, it was observed that the sound speed appears largely density-independent \cite{Zdunik:2012dj}. 
For the covariant nlNJL model such an observation has not yet been made.
Therefore, in the present work it is for the first time considered how well the covariant nlNJL model in the 2SC phase can be approximated by the quark matter EoS with constant speed of sound, given by Eq.~\eqref{eq:CSS_EoS}.

The CSS parameterization has been used recently for describing the hybrid stars in \cite{Somasundaram:2021ljr, Drischler:2020fvz}. 
It has been phenomenologically shown that PSR J0740+6620 could be described very well as a hybrid star with quark core in \cite{Somasundaram:2021ljr}. In that work the value of $c_s^2$ was chosen to be $1/3, 2/3,$ or $3/3$ and it was apparent that a value between 2/3 and 3/3 (for instance $c_s^2 = 1/2$) could best describe the radius for hybrid star in agreement with the recent NICER results. 
In \cite{Drischler:2020fvz}, they have shown that a $2.6 \textmd{M}_{\odot}$ NS would
require $c_s^2 \geq 0.55 - 0.6 $ in the inner core, and a $c_s^2 \approx 0.45 - 0.6$ would potentially be compatible with the NICER results for the radius of NS which also imply a high value of maximum mass. 
{Not only in the core of high mass pulsars, but also for a very light NS like the central compact object in the supernova remnant HESS J1731-347 with a mass $\textmd{M} = 0.77^{+0.20}_{-0.17} \textmd{M}_{\odot}$ and radius
$\textmd{R} = 10.4^{+0.86}_{-0.78}$ km recently reported by \cite{doroshenko2022strangely}, because of the untypically small values of mass and radius, the composition is likely to be superconducting quark matter rather than just ordinary nucleonic matter. Therefore, the hypothesis of forming color superconducting quark matter already in the cores of NS at subsolar masses is supported from phenomenological point of view, thus worth further investigation.}

It would be a fantastic finding that the EoS of a nonlocal chiral quark model as one of the most
advanced microscopic models for color superconducting quark matter, 
{with the favorable parameters found by a Bayesian analysis}, can be described with high
accuracy by the simple constant sound speed EoS that is widely used in the phenomenology of
compact stars with quark matter cores. While in many phenomenological models the squared sound speed is assumed to be 
1/3 (conformal limit) or 1 (causal limit), we find in a systematic way that a value close to 1/2 describes color superconducting quark matter. 
{This result is in agreement with \cite{Contrera:2022tqh}, where a 3D form factor is considered.}
{Furthermore, varying the two parameters of the nlNJL model enables us to perform an 
investigation on the trains of special point in the M-R diagram of hybrid stars.}

In order to test our hypothesis, we perform a fit of the nlNJL EoS to the CSS one for the relevant range of the nlNJL parameters $(\eta_v,\eta_D)$ including a $\chi^2$ analysis
in Appendix \ref{app:CSS-map}.
In that Appendix \ref{app:CSS-map}, we also provide a set of equations for mapping the nlNJL parameters to the ones defining the CSS model.

\section{Results and discussion}
\label{sec:Results}

\begin{figure}[thb]
    \includegraphics[width=0.50\textwidth]{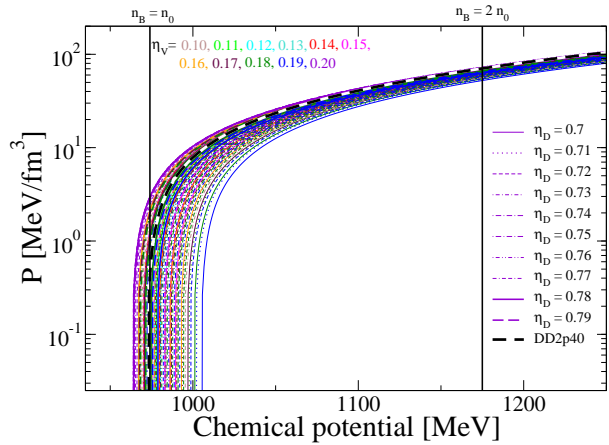}
    \caption{The pressure as a function of baryonic chemical potential for the Maxwell construction of hybrid stars over the whole range of $\eta_V$ and $\eta_D$. 
    Different colors correspond to different values of $\eta_V$ while different line styles correspond to different values of $\eta_D$.
    {The vertical dashed lines indicate the chemical potential at which the hadronic phase model (DD2p40) gives nuclear saturation density ($n_0$) and $2 n_0$}, respectively.}
    \label{fig:p-mu}
\end{figure}

In order to obtain realistic EoS models for hybrid neutron stars with quark matter cores, one has to add
a hadronic phase and construct the hadron-to-quark matter phase transition.
For the hadronic phase, we have chosen the relativistic density functional EoS "DD2p40", which captures the quark substructure effect of quark Pauli blocking among nucleons 
\cite{Blaschke:2020qrs} by a modified excluded nucleon volume procedure \cite{Typel:2016srf}, see the black solid line in Fig.~\ref{fig:p-mu}. 
For the phase transition construction we have chosen a standard Maxwell construction, where the critical pressure $P_c$ and the critical chemical potential $\mu_c$ of the first-order transition are obtained from the crossing of hadronic and quark matter lines in the  $P-\mu$ diagram, see Fig.~\ref{fig:p-mu}.
In that figure,  the vertical dashed lines indicate the chemical potential at nuclear saturation density ($n_0$) and at $2 n_0$. 
Applying a saturation density constraint, one would abandon all quark matter EoS for which the crossing with the hadronic EoS occurs to the left of this line.


\begin{figure}[thb]
    \includegraphics[width=0.5\textwidth]{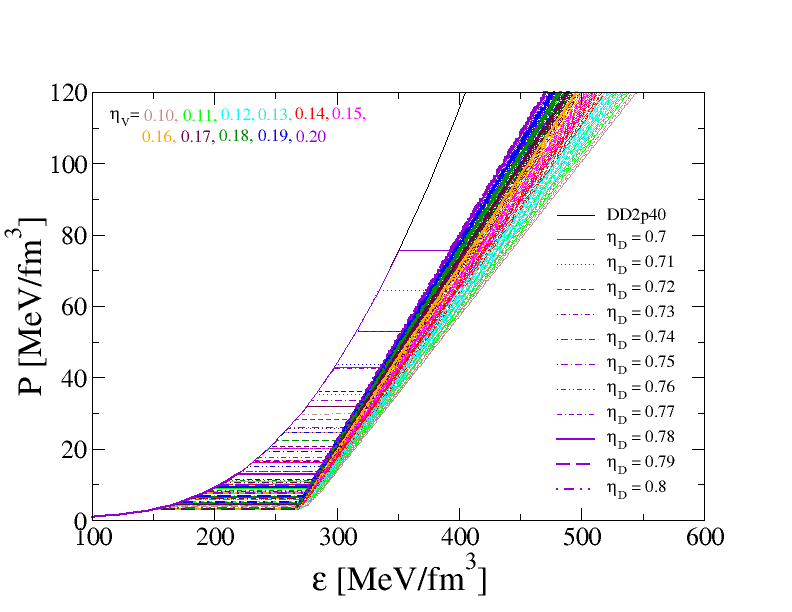}
    \caption{The pressure as a function of energy density for the Maxwell construction of hybrid stars over the whole range of $\eta_v$ and $\eta_D$. Colors and line styles correspond to variations of $\eta_V$ and $\eta_D$, respectively, as in Fig. \ref{fig:p-mu}.}
    \label{fig:p-e}
\end{figure}

{In Fig.~\ref{fig:p-e}, we show the resulting hybrid EoS in the 
plane pressure versus energy density $P(\varepsilon)$ for the whole domain of parameters which exhibit jumps in the energy density at the corresponding critical pressure which are characteristic for first-order phase transitions. In this figure, different values of $\eta_V$ are shown with different colors while different line styles correspond to different values of $\eta_D$.}

\begin{figure}[htb]
    \includegraphics[width=0.5\textwidth]{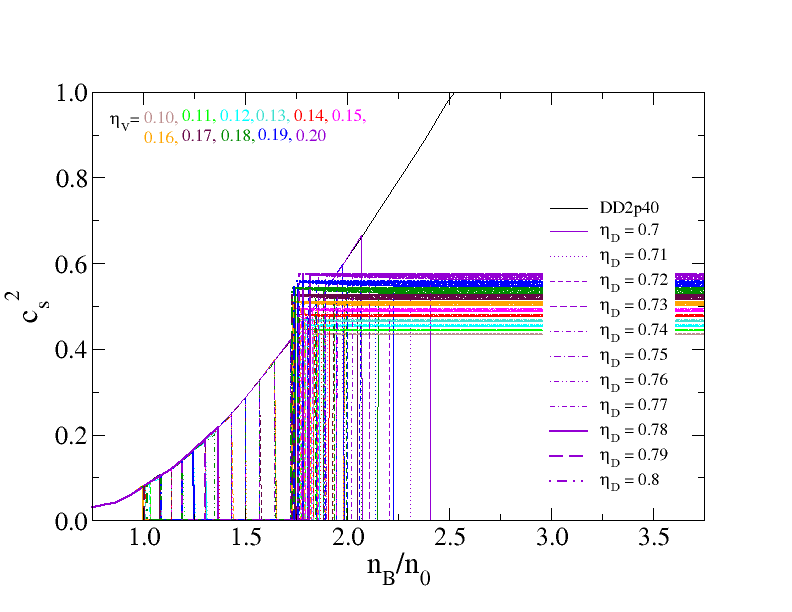}
    \caption{The squared speed of sound for the Maxwell construction of the hybrid EoS over the whole range of $\eta_V$ and $\eta_D$ parameters encoded by different line colors and styles, respectively, as in 
    Fig. \ref{fig:p-e}.}
    \label{fig:cs2-n}
\end{figure}

In Fig.~\ref{fig:cs2-n} we show the squared sound speed
$c_s^2=dP/d\varepsilon$ as a function of the baryon density 
in unit of the saturation density $n_0=0.15$ fm$^{-3}$ which 
vanishes in the region of the (energy) density jump because of the vanishing gradient of the pressure.
One clearly identifies the CSS quark matter phases in this figure with values of the squared sound speed in the range
$0.45 < c_s^2 < 0.54$ for the whole domain of parameters, see Tab.~\ref{tab:Fit_34}.

\begin{figure}[htb]
    \includegraphics[width=0.5\textwidth]{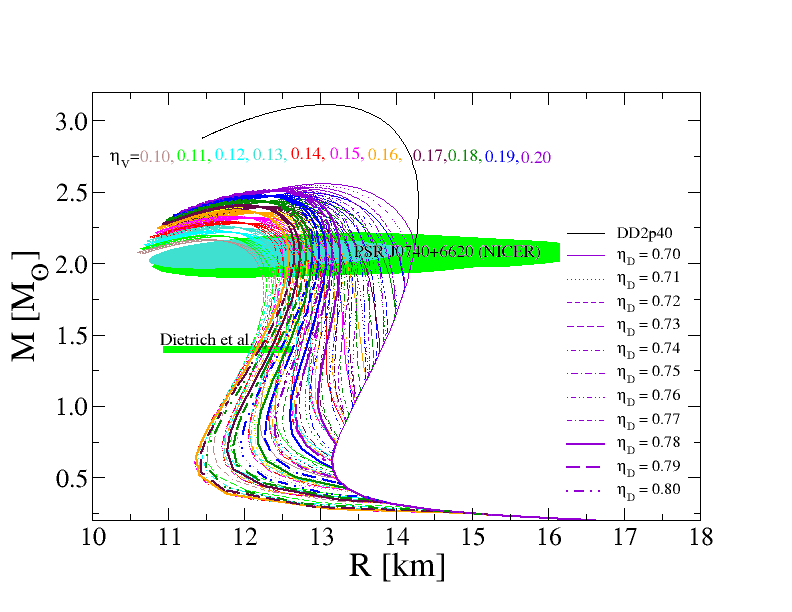}
    \caption{Mass-radius relations for the Maxwell construction of the hybrid EoS over the whole range of $\eta_V$ and $\eta_D$. Their encoding with different line colors and styles is the same as in Fig. \ref{fig:p-e}. For a comparison, the
1$\sigma$ mass-radius constraints from the NICER analysis of observations of the massive pulsar PSR J0740+6620 \cite{fonseca2021refined} are indicated as blue \cite{riley2021nicer} and green \cite{miller2021radius} regions. Additionally, the green bar marks the radius constraint for a 1.4 solar mass neutron star from the joint analysis of the gravitational-wave signal
GW170817 with its electromagnetic counterparts at 90\% confidence \cite{dietrich2020multimessenger}.}
    \label{fig:M-R}
\end{figure}

{The M-R plot is presented in Fig. \ref{fig:M-R}. We show
the hybrid solutions for the Maxwell construction together with the one for the purely hadronic model DD2p40. The $\eta_V$ and $\eta_D$ parameters were considered running over the whole range. 
We note that an interesting phenomenon becomes apparent in Fig. \ref{fig:M-R}: 
The sets of $M-R$ curves for a fixed value of $\eta_V$ (shown with the same color) and varying $\eta_D$ get collimated in focal points, the so-called "special points (SP)", that form a "train" with coordinates 
$(R_{\rm SP},M_{\rm SP})$, well described by a straight line in the $M-R$ diagram. 
This line plays an important role for NS phenomenology since each special point is closely related to the maximum mass $M_{\rm max}$ of a given hybrid EoS by the relation  \cite{Blaschke:2020vuy,Ivanytskyi:2022oxv}
\begin{equation}
M_{\rm max}=M_{\rm SP} + \delta (M_{\rm SP} - M_{\rm onset})~,
\end{equation}
where $\delta$ is a small, positive parameter depending on the class of quark matter EoS and $M_{\rm onset}$ is the mass of the NS for which the onset of deconfinement occurs in its center. 
More details are given in Appendix \ref{app:sp}.

Finally, in Fig. \ref{fig:Lambda-M} we show the tidal deformability  as a function of 
$\textmd{M}/\textmd{M}_{\odot}$ including the $\Lambda_{1.4}$ constraint from the low-spin prior analysis from GW170817 \cite{abbott2018gw170817}.}

\begin{figure}[htb]
    \includegraphics[width=0.5\textwidth]{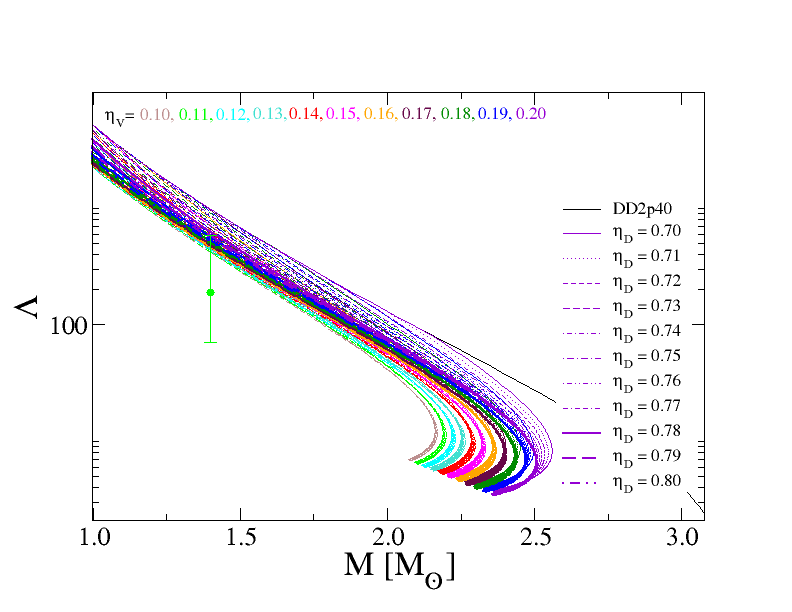}
    \caption{Dimensionless tidal deformability $\Lambda$ as a function of the star mass for the Maxwell construction of hybrid EoS over the whole range of $\eta_V$ and $\eta_D$. The description of different line colors and different line styles are the same as Fig. \ref{fig:p-e}. The green vertical line shows the $\Lambda_{1.4}$ constraint from the low-spin prior analysis of GW170817 \cite{abbott2018gw170817}.}
    \label{fig:Lambda-M}
\end{figure}

\section{Bayesian Analysis}\label{sec:Bayasian}

For the Bayesian analysis we have applied two constraints from neutron stars observations. The first one is the mass-radius constrains from PSR J0740+6620 (the neutron star component of a binary system with a white dwarf). 
Its gravitational mass has been measured with the relativistic Shapiro time delay effect based on data from the 100-m Green Bank Telescope and the Canadian Hydrogen Intensity Mapping Experiment telescope and it is ${2.08}_{-0.07}^{+0.07}$~M$_{\odot}$ (68.3\% credibility) \cite{Cromartie:2019kug, fonseca2021refined}. 
Its radius has been estimated with fits of rotating hot spot patterns to data from Neutron Star Interior Composition Explorer (NICER) and X-ray Multi-Mirror (XMM-Newton) X-ray observations and it is ${13.7}_{-1.5}^{+2.6}$ km (68\%) \cite{miller2021radius}.

The second one is the radius constraint for a 1.4~M$_{\odot}$ mass neutron star, which is ${11.75}^{+0.86}_{-0.81}$~km at 90\% confidence. This was estimated within a joint analysis of the gravitational-wave event GW170817 with its electromagnetic counterparts AT2017gfo and GRB170817A, and the gravitational-wave event GW190425, both originating from neutron-star mergers \cite{dietrich2020multimessenger}.

Additionally, we have introduced a lower limit for the density of deconfinement in neutron star matter as an additional constraint. We have chosen the saturation density $n_0$ as this lower limit. 
It was realized as a cumulative normal distribution with width $\sigma = 0.1~n_0$. 
In order to avoid the onset of deconfinement to occur below the saturation density $n_0$ beyond the $5~\sigma$ level, we chose the expectation value of the distribution to be $1.5~n_0$.

\begin{figure}[h]
    \includegraphics[width=0.40\textwidth]{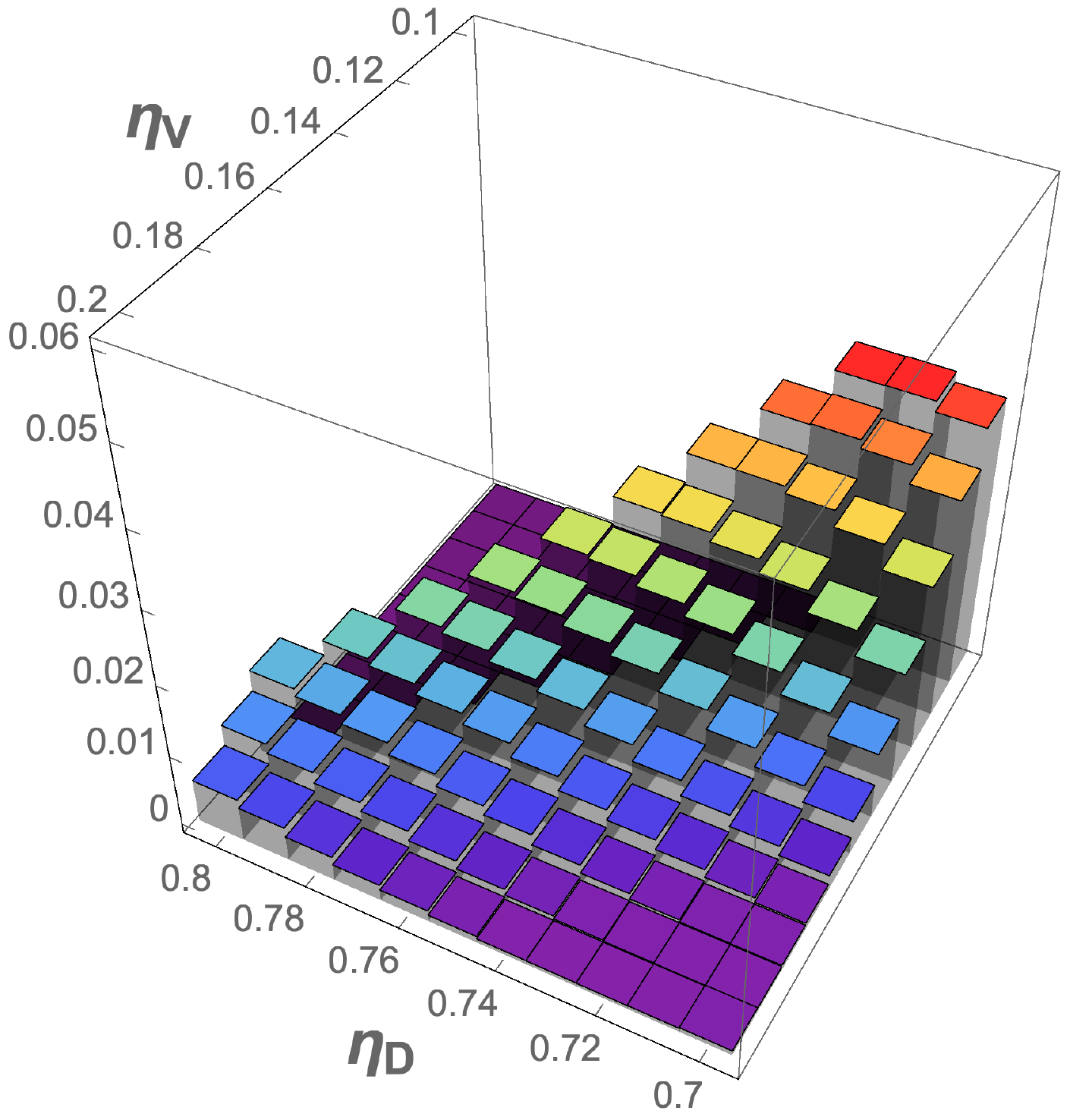}\\
    \includegraphics[width=0.40\textwidth]{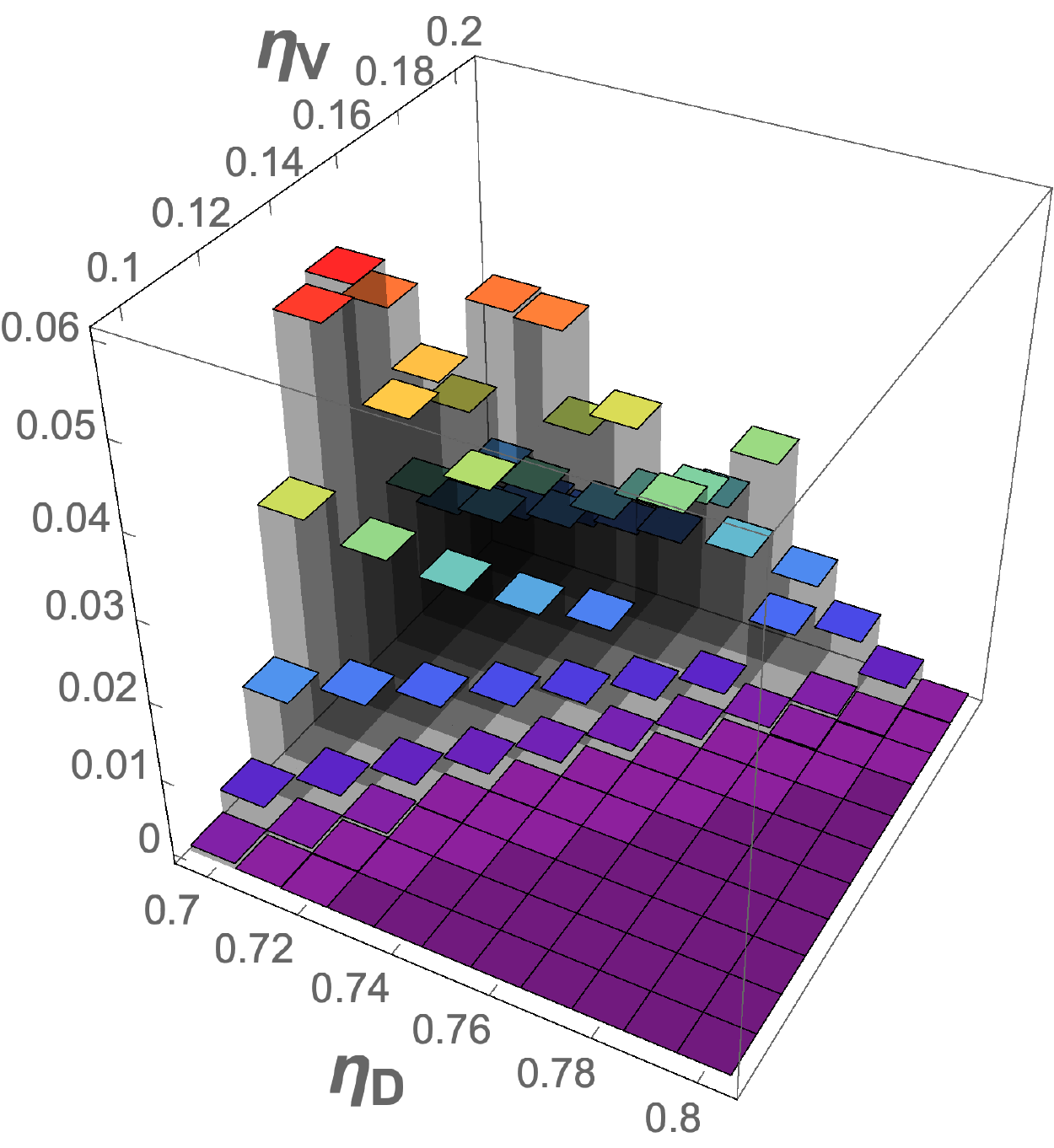}
    \caption{Bayesian analysis using the mass-radius measurement for PSR J0740+6620 and the radius constraint for a 1.4~M$_{\odot}$ mass neutron star for the class of hybrid EoS
obtained within a Maxwell construction between DD2p40 and nlNJL in the two-dimensional EoS parameter plane, $\eta_V$
and $\eta_D$. In upper panel, there is no constraint on the transition onset to quark matter while in lower panel it has been assumed that the transition can occur only after saturation density.}
    \label{fig:Bayasian}
\end{figure}

{From Fig. \ref{fig:Bayasian}, one could see that without enforcing the onset of the  transition  to quark matter to occur after saturation density, the probability of the parameters diagonally increases with decreasing both values of $\eta_V$ and $\eta_D$. But employing the onset density constraint has a remarkable effect on the results of Bayesian analysis.
Making the onset density of the first-order phase transition to occur after the saturation density is a physical constraint since we don't expect deconfined quark matter below saturation density of nuclear matter which is closely related to the interior density of heavy atomic nuclei.
Employing this constraint results in a favorable parameter range for the nlNJL model to be $\eta_V$ = 0.12-0.14 and $\eta_D$ = 0.70-0.71. These values of parameters are the best compatible ones with the employed observational constraints in the Bayesian analysis of this work.
In this sense, we have "measured" the values of the a priori unknown parameters $\eta_V$ and $\eta_D$ in the effective low-energy QCD Lagrangian using neutron star phenomenology!
}

\section{Conclusion} 
\label{sec:conslusion}
{In this work, we have found the favorable values for the vector meson coupling and the diquark coupling in the microscopic nlNJL model by performing a Bayesian analysis when a first-order phase transition from hadronic matter described within DD2p40 to color superconducting quark matter is constructed.}

The correspondence between the covariant nlNJL model and the CSS model EoS was considered. 
We have performed a mapping between the parameters of these two models in a decent range with a $\chi^2$ value that qualifies an excellent fit.
The finding of this equivalence allows to employ the simpler CSS approach instead of the covariant nlNJL model when a hybrid star EoS with color superconducting quark matter shall be constructed. 
The functional fit provided in this work allows to interpret the parameters of the CSS model that are favorable for explaining NS phenomenology in terms of the unknown coupling constants of the effective low-energy QCD Lagrangian. 
Therefore, we could finally confirm that the covariant nlNJL is well fitted to the CSS parameterization with high accuracy. 
While these microphysical parameters (the diquark coupling and the vector meson coupling) are in this work allowed to vary in the range of 0.7 $< \eta_D <$ 0.8 and 0.11 $< \eta_V <$ 0.18, respectively, it is possible to extend the range of applicability of the CSS parameter fit also beyond these ranges. 
Utilizing a Bayesian analysis with neutron star phenomenology, we could determine the most probably values for these a priori unknown parameters of the model Lagrangian for nonperturbative low-energy QCD! 

{The equivalence of this covariant nlNJL model to a CSS model 
parametrization has an interesting consequence which is discussed in greater detail in the Appendix \ref{app:sp}.
We have shown that simultaneously changing the two parameters ($\eta_V$ and $\eta_D$) and understanding this as changing the three parameters of the CSS model (A, B, and $c_s^2$) results in the appearance of trains of special points instead of one special point which emerges when only B is changed. 
We have shown that a simultaneous variation of of vector and diquark coupling by fixed steps  $\delta \eta_V$ and $\delta \eta_D$  while keeping the ratio of variation $\xi=\delta \eta_V/\delta \eta_D$ fixed defines the lines (trains) along which the special points are located. 
We have performed linear fits describing the position of the trains of special points in the mass-radius diagram and showed that they remain unchanged when varying the hadronic matter EoS from DD2p40 to DD2p00. 
The line corresponding to $\xi=0$ plays a special role for the phenomenology of compact stars. 
This line parametrizes the train of special points with the highest slope in the M-R diagram, a proxy for the lower limit on the maximum mass of hybrid NS as a function of $\eta_V$, i.e. when varying the stiffness of quark matter. This is an important and original achievement of the present study.}

Our studies can be used not only in order to explore the phase transition between hadronic matter and color superconducting quark matter in cold NS, but it can also be extended to finite temperatures when it will be useful in simulations of core collapse supernovae \cite{Fischer:2017lag}
and neutron star mergers \cite{Bauswein:2018bma,Bauswein:2020aag,Blacker:2020nlq} involving quark matter deconfinement.

\begin{acknowledgments}
SA acknowledges the COST Action CA16214 “PHAROS” for the support of this project through the Short Term Scientific Mission (STSM), and appreciates the hospitality of the Physics Department of Faculty of Science at University of Zagreb for having her as a guest scientist.
This work has been supported in part by the Polish National Science Centre (NCN) under grant No. 2019/33/B/ST9/03059 (AA,DB,MS,SA)
and Consejo Nacional de Investigaciones Cient\'ificas y T\'ecnicas, under Grant No.~PIP17-700 (AGG).
\end{acknowledgments}

\appendix
\section{Mapping procedure}
\label{app:CSS-map}

In order to obtain a functional relation between the parameters of the two approaches, we considered 34 EoS based on nlNJL model (for different values of $\eta_D$ and $\eta_V$ parameters) to which we fit the CSS EoS (\ref{eq:CSS_EoS}) in order to obtain the values of $A$, $B$ and $\beta$. In these EoS fits, the parameter $B$ is expressed through the values of $A$ and $\beta$ as  
\begin{equation}
    B = A \left( \frac{\mu(P=0)}{\mu_x}\right)^{1+\beta},
\end{equation}
with the value of chemical potential for which the pressure is equal to zero, given by the nlNJL EoS.

\begin{figure}[b!]
    \centering
    \includegraphics[width=\linewidth]{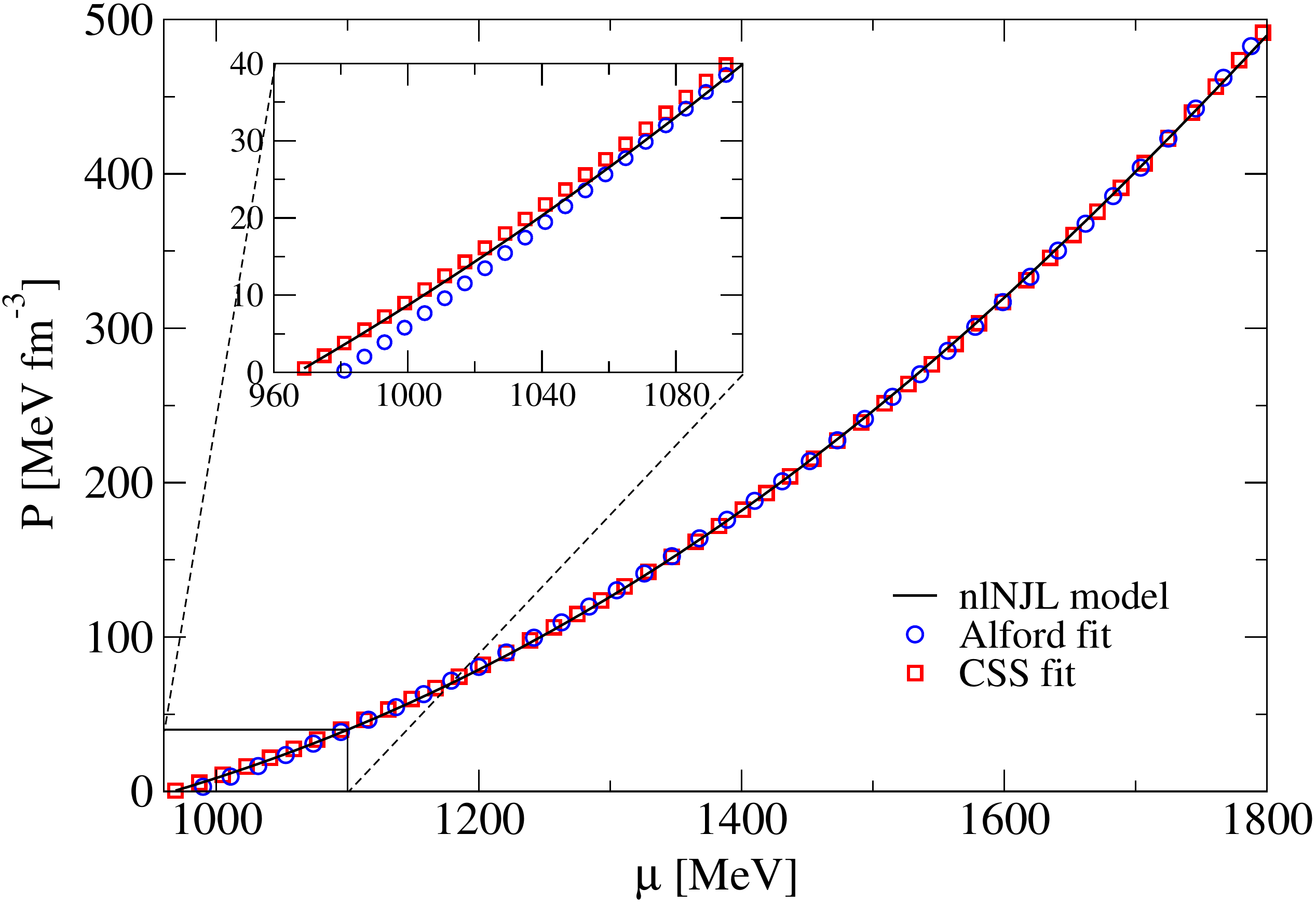}
    \caption{EoS of the original nlNJL model for the parameter set $(\eta_D,\eta_V)=(0.75,0.14)$ compared to its fit by the phenomenological form \eqref{eq:quartic} (Alford fit, blue circles) with the parameters $a_4=0.221$, $a_2=-(307.085~\textmd{MeV})^2$ and $B_{\textrm{eff}}^{1/4}=175.658$ MeV 
    as well as the fit by the CSS pressure of Eq.~\eqref{eq:CSS_EoS} (red squares)
with the parameters $A$ = 94.875, $c_s^2$ = 0.481 and $B$ = 85.614, see Tab.~\ref{tab:Fit_34}. }
    \label{fig:gracefit}
\end{figure}

One example of the CSS fit to the nlNJL EoS (with $\eta_D=0.75$ and $\eta_V=0.14$) is given in Fig.~\ref{fig:gracefit}. For comparison, fitting the parameters of the phenomenological EoS \eqref{eq:quartic} that was introduced by Alford et al. in \cite{Alford:2004pf} we obtain $a_4=0.221$, $a_2=-(307.085~\textmd{MeV})^2$ and $B_{\textrm{eff}}^{1/4}=175.658$ MeV. This "Alford fit" is also shown in Fig.~\ref{fig:gracefit}. 
It is worth mentioning that the value of $a_4\approx 0$, which is obtained by the fit, entails that the correction relative to the ideal gas limit is close to 100\% and thus not compatible with its value of $c=2\alpha_s/\pi\approx 30\%$ in perturbation theory to $O(\alpha_s)$ \cite{Fraga:2001id}. This can be seen as an hint to the nonperturbative nature of the quark matter EoS in this domain of low chemical potentials in the vicinity of the hadronization transition. 

In the inserted plot we further examine the low chemical potential region where the most significant difference between the two fits is found. The CSS follows the nlNJL behaviour more closely, which will be also manifested in the mass-radius curves discussed toward the end of the chapter. For the remaining chemical potentials, the two fits present the same quality in reproducing the nlNJL EoS. The quality of the fits is estimated with the $\chi^2$ value, defined as 
\begin{equation}
    \chi^2 = \sum_i^N \frac{(P_{\textrm{nlNJL}}(\mu_i) - P_{\textrm{fit}}(\mu_i))^2}{\sigma_i^2} ~,
\end{equation}
where $N$ is the number of points for the chemical potential and $\sigma$ is the standard deviation of the nlNJL model in question defined as
$\sigma^2 = \frac{1}{N}\sum_i^N (P_{\textrm{nlNJL,} i} - \overline{P_i})^2$, where $\overline{P_i}$ is the mean value of the nlNJL model pressures, $\overline{P} = \frac{1}{N} \sum_i^N P_i$. The $P_{\textrm{nlNJL}}(\mu_i)$ and $P_{\textrm{fit}}(\mu_i)$ are the values of pressure in nlNJL model and for the fit in each point of chemical potential $\mu_i$. The $\chi^2$ value for CSS fit is 0.031 while it is 0.038 for Alford's fit.

From the CSS fit, the values of $A$ and $\beta$
parameters are obtained, from which $B$ parameter and squared speed of sound $c_s^2$ are calculated. These values are given in Table~\ref{tab:Fit_34} together with the $\chi^2$ values for the 34 different nlNJL EoSs giving the total of 34 data points. From these values the functional form, between nlNJL parameter space and the CSS one, is to be found. 

\begin{table} [t]
\caption{\label{tab:Fit_34}
 The values of $A$, $B$ and c$_s^2$ calculated from the CSS EoS fit to the given nlNJL model defined by the values of $\eta_D$ and $\eta_V$.}
\begin{ruledtabular}
\begin{tabular}{lccccc}
$\eta_D$ & $\eta_V$ & $A$ &  $c_s^2$ & $B$ & $\chi^2$ \\
&   & [MeV/fm$^{3}$] & [$c^2$] & [MeV/fm$^{3}$] & \\
\hline
0.70&	0.15&	91.484&	0.488&	87.209&	0.039\\ 
0.71&	0.12&	91.053&	0.456&	83.425&	0.022\\
0.71&	0.14&	91.649&	0.476&	85.815&	0.032\\
0.71&	0.16&	92.963&	0.502&	89.021&	0.047\\
0.71&	0.18&	94.481&	0.532&	92.214&	0.075\\
0.72&	0.13&	92.132&	0.467&	84.592&	0.026\\ 
0.72&	0.15&	92.954&	0.490&	87.209&	0.038\\ 
0.72&	0.17&	94.366&	0.517&	90.408&	0.058\\
0.73&	0.12&	92.612&	0.457&	83.280&	0.021\\
0.73&	0.14&	93.190&	0.478&	85.658&	0.031\\
0.73&	0.16&	94.170&	0.503&	88.385&	0.048\\
0.73&	0.18&	96.211&	0.535&	92.290&	0.073\\
0.74&	0.11&	93.236&	0.449&	82.095&	0.017\\
0.74&	0.13&	93.563&	0.468&	84.217&	0.026\\
0.74&	0.15&	94.410&	0.491&	86.884&	0.039\\
0.74&	0.17&	95.780&	0.519&	90.011&	0.061\\
0.75&	0.12&	94.000&	0.461&	82.899&	0.044\\
0.75&	0.14&	94.875&	0.481&	85.614&	0.031\\
0.75&	0.16&	95.894&	0.506&	88.391&	0.056\\
0.75&	0.18&	97.934&	0.538&	92.249&	0.078\\
0.76&	0.13&	95.235&	0.470&	84.101&	0.027\\
0.76&	0.15&	96.153&	0.494&	86.873&	0.039\\
0.76&	0.17&	97.660&	0.522&	90.172&	0.063\\
0.77&	0.12&	95.556&	0.461&	82.437&	0.021\\
0.77&	0.14&	96.433&	0.483&	85.287&	0.032\\
0.77&	0.16&	97.770&	0.509&	88.512&	0.074\\
0.77&	0.18&	99.685&	0.541&	92.155&	0.085\\
0.78&	0.15&	97.485&	0.495&	86.179&	0.042\\
0.78&	0.17&	99.340&	0.525&	90.034&	0.065\\
0.79&	0.12&	97.604&	0.464&	82.718&	0.020\\
0.79&	0.14&	97.912&	0.484&	84.755&	0.033\\
0.79&	0.16&	99.216&	0.511&	87.929&	0.053\\
0.79&	0.18&	100.878&0.541&	91.415&	0.084\\
0.80&	0.17&	101.116&0.528&	89.766&	0.070
\end{tabular}
\end{ruledtabular}
\end{table}

\begin{figure*}[t!]
    \centering
     \includegraphics[scale=0.7]{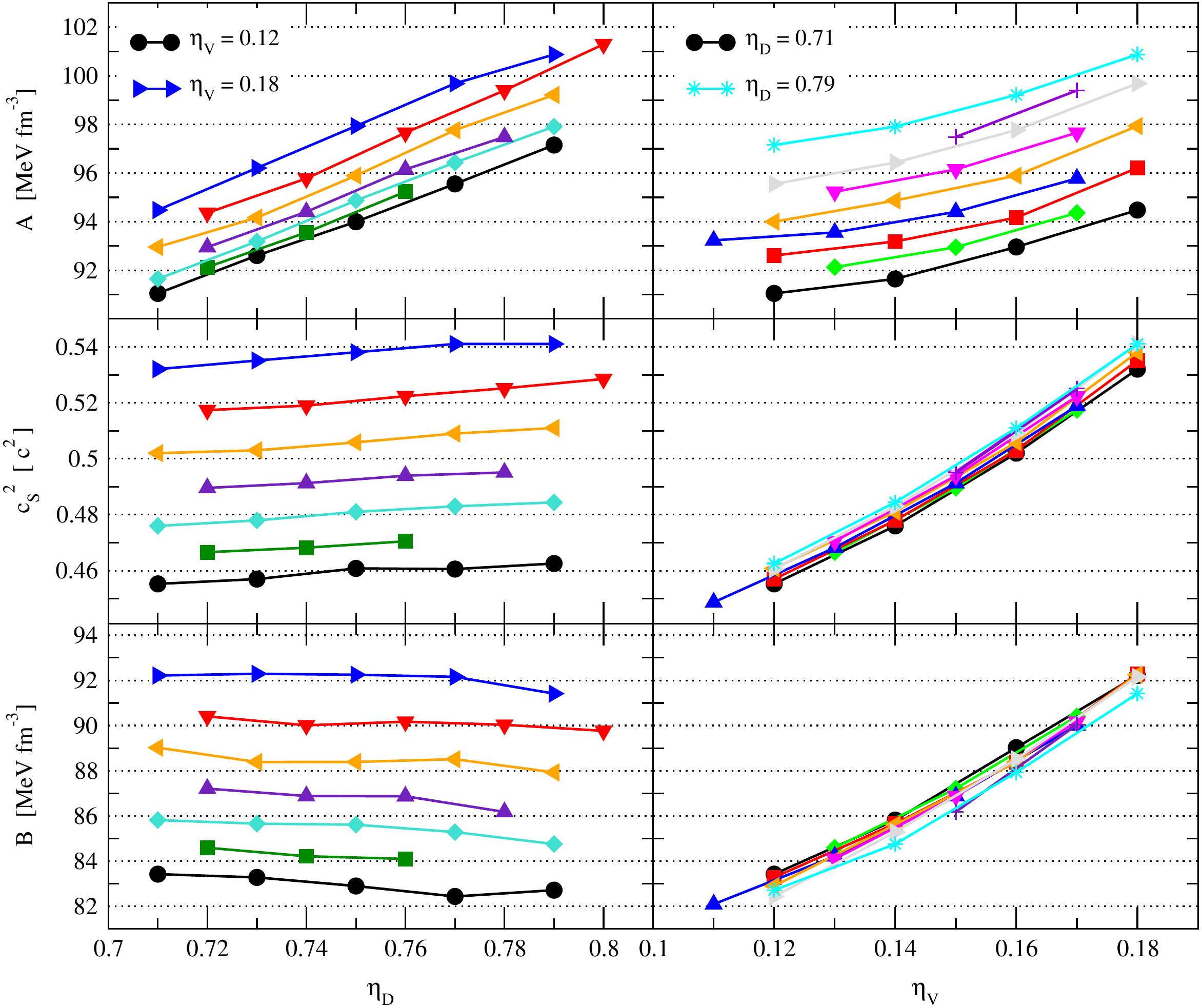}
    \caption{The change of CSS parameters ($A$, $c_s^2$ and $B$) with the increase of $\eta_D$ for different values of $\eta_V$ (from 0.12 to 0.18 in steps of 0.01) on the left panels, and with the increase of $\eta_V$ for different values of $\eta_D$ (from 0.71 to 0.78 in steps of 0.01) on the right panels.}
    \label{fig:Param_etaDetaV}
\end{figure*}

In order to find the functional dependence between the two parameter spaces, we analyze the behaviour of the CSS model parameters for different values of $\eta_D$ and $\eta_V$, as given in Figure \ref{fig:Param_etaDetaV}. 
The parameter $A$ shows the strongest dependence on both nlNJL model parameters, while $c_s^2$ and $B$ are almost independent on $\eta_D$ and linearly depending on $\eta_V$, respectively. From these dependencies, the simplest functional form between two parameter spaces can be assumed: the variation of CSS parameters is linear with the change of $\eta_D$, while it is quadratic with the change of $\eta_V$. Thus, we write the following equations
\begin{eqnarray}
\label{eq:A}
    A &=& a_1 \eta_D + b_1 \eta_V^2 + c_1 \\
\label{eq:cs2}
    c_s^2 &=& a_2 \eta_D + b_2 \eta_V^2 + c_2\\
    B &=& a_3 \eta_D + b_3 \eta_V^2 + c_3 ,
\label{eq:B}
\end{eqnarray}
where the coefficients $a_i$, $b_i$ and $c_i$ (with $i = 1, 2, 3$) are obtained 
through a two-parameter ($\eta_D$, $\eta_V$) fitting procedure of each of the CSS parameters ($A$, $B$, c$_s^2$) and are given in Table~\ref{tab:Coefficients}.

\begin{table}[thb]
\caption{ The values of $a_i$, $b_i$ and $c_i$ coefficients ($i=1,2,3$) for 
the mapping between the ($\eta_D,\eta_V$) and ($A$, c$_s^2$, $B$) parameter spaces given by 
Eqs.~\eqref{eq:A}-\eqref{eq:B}.
}
\label{tab:Coefficients}
\begin{ruledtabular}
\begin{tabular}{cccccc}
i & parameter& unit &  $a_i$ & $b_i$ & $c_i$  \\
\hline
1 & $A$ & MeV\, fm$^{-3}$   & 80.66330 & 199.80900 & 30.57520 \\ 
2 & $c_s^2$ & $c^2$ & 0.11205 & 4.31830 & 0.31244  \\
3 & $B $ & MeV\, fm$^{-3}$   & -10.42990 & 502.99800 & 83.46230\\
\end{tabular}
\end{ruledtabular}
\end{table}

\begin{figure}[t!]
    \centering
    \includegraphics[width=\linewidth]{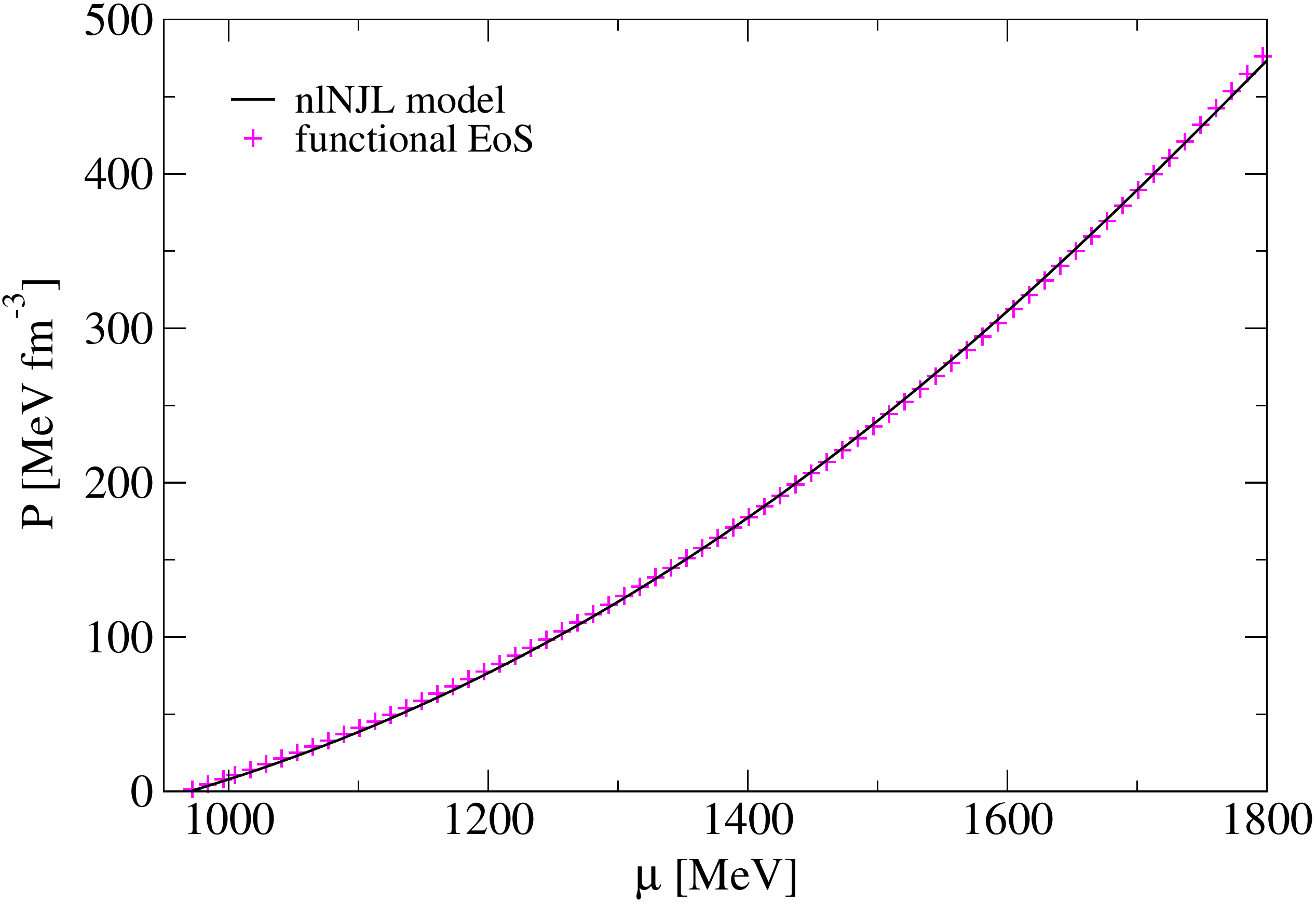}
    \caption{
    Comparison of the EoS for the original nlNJL for $\eta_D=0.75$ and $\eta_V=0.15$ (black solid line) 
    with its CSS representation (magenta pluses) calculated from
    Eq.~\eqref{eq:CSS_EoS} where the parameters $A$ = 95.568 $c_s^2$ = 0.494 and  $B$ = 86.957
    are determined using the functions \eqref{eq:A}-\eqref{eq:B} with the coefficients from Tab.~\ref{tab:Coefficients}.
    The $\chi^2$ value of the fit is 0.049.
    }
    \label{fig:075017}
\end{figure}

To check the quality of our method, we choose one original nlNJL EoS that was not included in the initial fitting data set (e.g. $\eta_D=0.75$ and $\eta_V=0.15$) and calculate CSS EoS using Eq.~\eqref{eq:CSS_EoS} with the parameter values obtained through Eqs.~\eqref{eq:A}-\eqref{eq:B}. The EoS comparison is given in Fig.~\ref{fig:075017}. It is worth noticing that the $M-R$ curve for our example EoS
with the parameter set $(\eta_D,\eta_V)=(0.75,0.15)$ from the middle of the parameter range crosses the revised mass value $2.08~\textmd{M}_\odot$ for PSR J0740+6620 \cite{fonseca2021refined} at the radius $R=12.5$ km, which accidentally is in the middle of the overlap region $12.2 {\rm ~km}< R <13.7$ km of the 1$\sigma$ NICER radius measurements from the two analysis teams (Riley et al. and Miller et al.) that have recently been reported \cite{NICERweb}.

\begin{figure}[!htb]
    \includegraphics[scale=0.7]{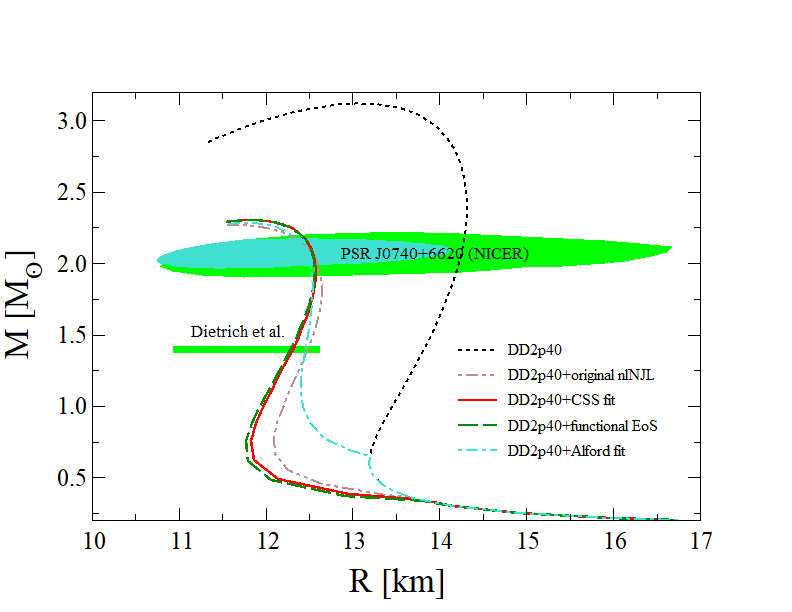}
\caption{
The M-R sequences for hybrid star EoS obtained by a Maxwell construction for a DD2p40 EoS with the original nlNJL model with $\eta_D = 0.75$, $\eta_V = 0.14$, its CSS fit, the functional mapping and the Alford fit.}
\label{fig:massradius}
\end{figure}

We have performed the fitting procedure considering the Alford parameterization as well. The results show that the CSS parameterization works better for fitting the nlNJL model. For comparison, the mass-radius ($M-R$) curves for the hybrid stars constructed using nlNJL model and two different versions of fitted EoS are shown in Fig.~\ref{fig:massradius}. For the hadronic EoS we have employed the relativistic density-functional (RDF) model based on the DD2 parameterization \cite{Typel:2009sy} with excluded volume effects \cite{Typel:2016srf}. The DD2 model with excluded
volume is mainly used in the context of an early deconfinement onset with large latent heat. 

The results for the $M-R$ curves sufficiently demonstrate the applicability of the mapping that has been performed. It is clear that the $M-R$ curves for the CSS fit and its functional version are sufficiently close for these two parameterization to be considered equivalent. Furthermore, their difference to the $M-R$ curve for the original nlNJL EoS amounts to maximally $200$ m in radius. This difference is well visible but still much smaller than the design accuracy of observational radius measurements from NICER, which even has not yet been reached by the first 
NICER radius measurements. Therefore, our work provides a sufficiently precise tool for NS phenomenology. Moreover, the comparison with Alford fit shows that the radius of the NS is really sensitive to the EoS. While both fits are very close according to Fig.~\ref{fig:gracefit}, the CSS parameterization performs better in the $M-R$ curve, in particular close to the onset of the phase transition at low densities. From our results, we can see that the Alford fit may be useful close to the maximum mass.

We conclude that for the nlNJL EoS parameterization within the ranges 0.7 $< \eta_D <$ 0.8 and 0.11 $< \eta_V <$ 0.18, the CSS approach can mimic the behaviour of the nlNJL EoS to high accuracy. The range of $\eta_D$ values between 0.7 and 0.8 is 
covering the value for the Fierz transformation of a one-gluon exchange interaction ($\eta_D = 3/4$) and entails that the quark matter is in the color superconducting phase. The CSS parameters for the nlNJL EoSs with high values of $\eta_V >$ 0.18 are showing deviations from the general behaviour fitted by Eqs.~\eqref{eq:A}-\eqref{eq:B}. But since for these $\eta_V$-values there is a causality  violation at large chemical potentials, the CSS form of the EoS fitted at low densities can be used to extrapolate the EoS above a certain density. A lower limit on the values of the $\eta_V$ parameter 
is not explored here.

\section{"Trains" of special points}
\label{app:sp}
\begin{figure*}[ht]
    \includegraphics[width=1.0\textwidth]{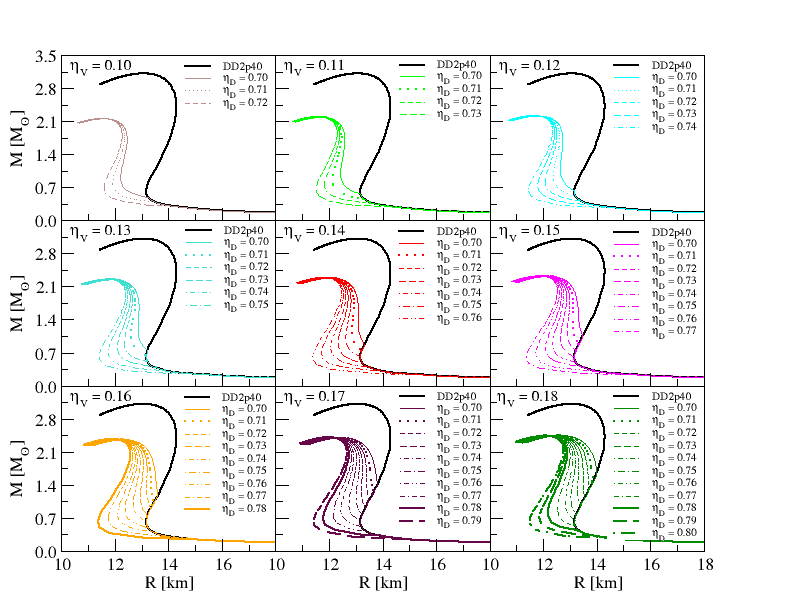}
    \caption{Mass-radius relation for all hybrid stars obtained by a Maxwell construction. The value of $\eta_V$ is taken to be fixed for each panel while the value of $\eta_D$ is varied.}
    \label{fig:M-R_matrix}
\end{figure*}

\begin{figure}[ht]
    \includegraphics[width=0.5\textwidth]{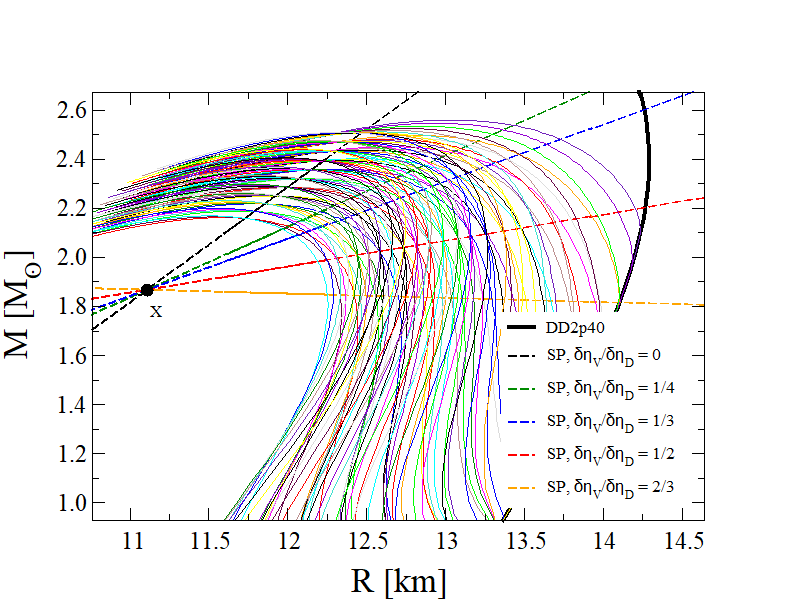}
    \caption{Mass-radius relation for all hybrid stars obtained by a Maxwell construction. Lines connect special points with a fixed slope $\xi=\delta \eta_V/(\delta \eta_D)$ of simultaneous variation of the Lagrangian parameters and meet in one point denoted as X. }
    \label{fig:M-R_lines}
\end{figure}

\begin{figure}[ht]
    \includegraphics[width=0.5\textwidth]{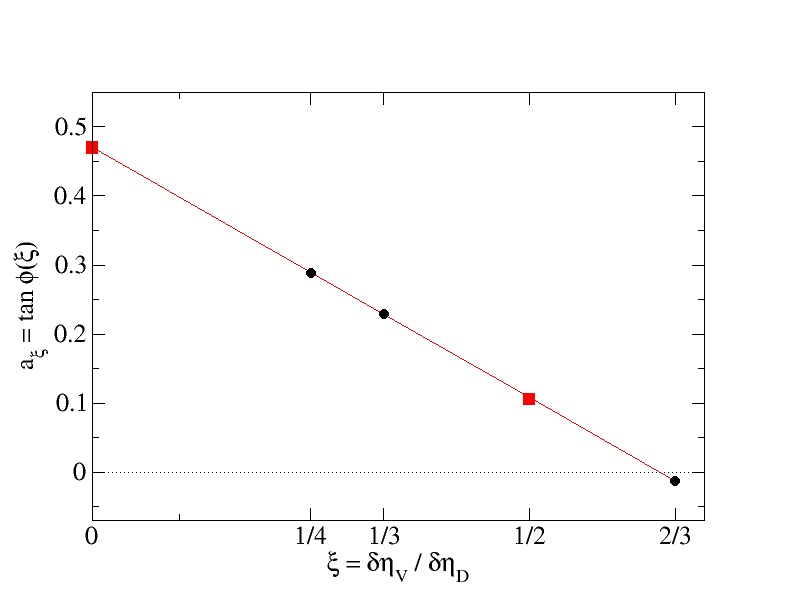}
    \caption{Dependence of the slope of the lines connecting special points on the slope $\xi=\delta \eta_V/(\delta \eta_D)$ of simultaneous variation of the Lagrangian parameters.}
    \label{fig:etaD-etaV}
\end{figure}

\begin{figure}[ht]
    \includegraphics[width=0.5\textwidth]{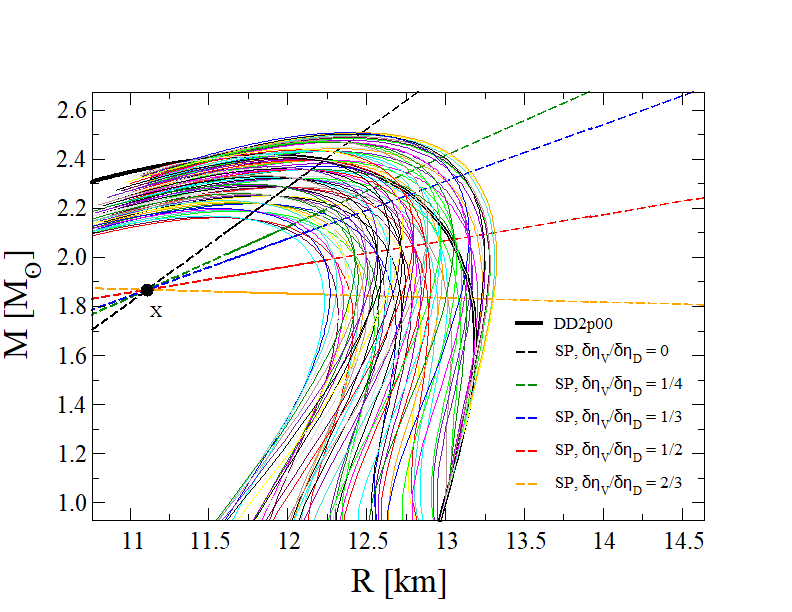}
    \caption{Same as Fig.~\ref{fig:M-R_lines} but for the softer hadronic EoS DD2p00. The lines connection special points and the point X where they all meet remain unchanged.}
    \label{fig:M-R_lines-2}
\end{figure}

In Fig. \ref{fig:M-R} one can see clearly that M-R curves corresponding to a fixed value of $\eta_v$ but different values of $\eta_D$ get collimated in a narrow region close to the maximum mass which has been dubbed "special point" (SP)
\cite{Yudin:2014mla}. Incrementing the value of $\eta_V$, a new SP is obtained, so that for our set of EoS a train of special points in the $M-R$ diagram emerges.
For better visibility of the effect, wee have selected a subset of 9 values for $\eta_V$ and show a 3x3 matrix of panels
with the corresponding $M-R$ diagrams in Fig.~\ref{fig:M-R_matrix}.
In Fig. \ref{fig:M-R_lines} we provide a detailed inspection of the location of the special points in the M-R diagram.
We find the following systematics which would hold for a simultaneous variation of the Lagrangian parameters of the quark matter EoS:
\begin{enumerate}
    \item Varying $\eta_V$ and $\eta_D$ simultaneously while keeping the ratio of variations fixed to $\xi=\delta \eta_V/\delta \eta_D$ defines a line $M^{(\xi)}(R)=a_\xi R + b_\xi $ in the M-R diagram along which special points are located.
    \item All these lines meet in one point denoted by "X" with the coordinates $(M_X,R_X)=(1.8663~\textmd{M}_\odot, 11.112~{\rm km})$.
    \item The slope of the lines follows a linear relationship 
    $a_\xi = \tan \phi(\xi) = \alpha - \beta \xi$ with $\alpha=0.47074$ and $\beta=0.7252$, see Fig.~\ref{fig:etaD-etaV}.
    \item The line of special points with largest slope corresponds to $\xi=0$, i.e. where for each special point on that line corresponds to a fixed $\eta_V$ so that $\delta \eta_V=0$.  This line is a lower limit approximation to the line of maximal masses as a function of $\eta_V$, i.e. the stiffness of quark matter EoS.
    \item We have repeated the analysis by replacing the DD2p40 with the DD2p00 EoS. The result shown in Fig. \ref{fig:M-R_lines-2} demonstrates the independence of the trains of special points from the hadronic EoS used in constructing the hybrid EoS model.  
\end{enumerate}

\section{Phenomenological EoS}
\label{app:P-EoS}

Besides the CSS model EoS, there is a phenomenological  formulation of the EoS of quark matter in use which has been introduced and motivated in Ref.~\cite{Alford:2004pf}. 
In that work, the quark matter EoS consists of the first three terms of a series in even powers of the quark chemical potential
\begin{equation}
\label{eq:quartic}
\Omega_{QM}=-3/4\pi^2 a_4 \mu^4 + 3/4\pi^2 a_2 \mu^2 + B_{\textrm{eff}}~,
\end{equation}
where $a_4$, $a_2$, and $B_{\textrm{eff}}$ are coefficients independent of $\mu$.

The quartic coefficient $a_4=1-c$ is well defined for an ideal massless gas for which c=0. 
Perturbative QCD corrections in lowest order, i.e. $O(\alpha_s)$ for massless quark matter lead to a reduction of $a_4$, e.g., accounted for by $c=0.3$ so that $a_4=0.7$, see also \cite{Fraga:2001id}. Since neutron stars "live" in the nonperturbative domain, it may be admissible to employ even larger values of $\alpha_s$ which lead to a further reduction of $a_4$ and a stiffening of the quark matter EoS  in the density domain of NS cores \cite{Blaschke:2021poc}. 
For this observation, the term quadratic in $\mu$ is important which arises from an expansion in the finite strange quark mass $m_s$ and the diquark pairing gap
$\Delta$, so that $a_2 = m_s^2-4\Delta^2$. 
Since in CFL quark matter both parameters are of the same order of about $100$ MeV, the coefficient $a_2$ becomes negative and dominated by the pairing gap. This corresponds to an effective  speed of sound exceeding the conformal limit $c_s^2=1/3$ which is approached, however, at large densities where the term $\propto \mu^4$ in the EoS (\ref{eq:quartic}) dominates \cite{Alford:2004pf,Blaschke:2021poc}. 

We show in Fig.~\ref{fig:massradius} that even for this simple parameterization (\ref{eq:quartic}) which we denote as "Alford fit", the resulting hybrid stars can have mass-radius relations very similar to those of the CSS fit and thus fulfill the maximum mass constraint \cite{fonseca2021refined}.

\bibliography{mapping}

\end{document}